\documentclass[manuscript,screen]{acmart}
\AtBeginDocument{%
  \providecommand\BibTeX{{%
    \normalfont B\kern-0.5em{\scshape i\kern-0.25em b}\kern-0.8em\TeX}}}

\setcopyright{acmcopyright}
\copyrightyear{2022}
\acmYear{2022}
\acmDOI{10.1145/3527927.3532789.}

\acmBooktitle{}
\acmPrice{15.00}
\acmISBN{978-1-4503-XXXX-X/18/06}

\usepackage{multirow, makecell}
\begin{document}


\title{Understanding User Perceptions, Collaborative Experience and User Engagement in Different Human-AI Interaction Designs for Co-Creative Systems}

\author{Jeba Rezwana}
\email{jrezwana@uncc.edu}
\author{Mary Lou Maher}
\email{m.maher@uncc.edu}
\affiliation{%
  \institution{University of North Carolina at Charlotte}
  \country{USA}
  \postcode{}
}

\renewcommand{\shortauthors}{}


\begin{abstract}
Human-AI co-creativity involves humans and AI collaborating on a shared creative product as partners. In a creative collaboration, communication is an essential component among collaborators. In many existing co-creative systems users can communicate with the AI, usually using buttons or sliders. Typically, the AI in co-creative systems cannot communicate back to humans, limiting their potential to be perceived as partners rather than just a tool. This paper presents a study with 38 participants to explore the impact of two interaction designs, with and without AI-to-human communication, on user engagement, collaborative experience and user perception of a co-creative AI. The study involves user interaction with two prototypes of a co-creative system that contributes sketches as design inspirations during a design task. The results show improved collaborative experience and user engagement with the system incorporating AI-to-human communication. Users perceive co-creative AI as more reliable, personal, and intelligent when the AI communicates to users. The findings can be used to design effective co-creative systems, and the insights can be transferred to other fields involving human-AI interaction and collaboration.
\end{abstract}




\maketitle

\section{Introduction}
Human-AI co-creativity, a subfield of computational creativity, involves humans and AI collaborating in a creative process as partners to produce creative artifacts, ideas or performances \cite{davis2013human}. Human-AI co-creativity is an important area of study given that AI is being used increasingly in collaborative spaces, including AI in collaborative music \cite{hoffman2010shimon, yee2016experience}, collaborative design \cite{karimi2018evaluating}, collaborative writing \cite{samuel2016design}, collaborative dance \cite{long2017designing}, or even in hospitals as a virtual nurse \cite{crowder2020human}. This growing field has prospects in many fields, especially education, industry and entertainment. Human-AI co-creativity has the potential to transform how people interact with an AI and user perception of an AI partner. Interaction is a fundamental property of co-creative systems as both the human and the AI actively participate and interact with each other, unlike autonomous creative systems that generate creative artifacts independently and creativity support tools (CST) that support human creativity \cite{kantosalo2019human}. Interaction design in a co-creative AI system has many challenges due to the open-ended nature of the interaction between the human and the AI \cite{davis2016empirically, kantosalo2014isolation}. Humans utilize many different creative strategies and reasoning processes throughout the creative process, and ideas and the creative product develop dynamically through time. This continual progression of ideas requires adaptability on the agent's part. AI ability alone does not ensure a positive collaborative experience of users with the AI \cite{louie2020novice} and interaction is more critical than algorithms where interaction with the users is essential \cite{wegner1997interaction}. Bown asserted that the success of a creative system's collaborative role should be further investigated through interaction design as interaction plays a key role in the creative process of co-creative systems \cite{bown2015player}.  

Communication is an essential part of interaction design in human-AI co-creativity \cite{bown2020speculative}. Stephen Sonnenburg demonstrated that communication is the driving force of collaborative creativity \cite{sonnenberg1991strategies}. Two-way communication between collaborators for providing feedback and sharing important information improves user engagement in human collaboration \cite{bryan2012identifying}. However, Rezwana and Maher showed in their recent work that many co-creative systems do not design for two-way communication between the user and the AI  \cite{rezwana2022designing}. They analyzed 92 co-creative systems and showed that all of the co-creative systems in their dataset use only human-to-AI communication, primarily with buttons, sliders, or other UI components, for users to communicate directly with the AI \cite{rezwana2022designing}. However, there is no channel for AI-to-human communication in most systems \cite{rezwana2022designing}. For example, Collabdraw \cite{fan2019collabdraw} is a co-creative system sketching environment where users draw with an AI. The user clicks a button to submit their artwork and indicate that their turn is complete. The AI in this system cannot communicate with users to provide information, suggestions, or feedback. While the AI algorithm is capable of providing intriguing contributions to the creative product, the interaction design does not focus on a successful human-AI collaboration. AI-to-human communication is an essential aspect of human-computer interaction and essential for a co-creative AI to be considered as a partner \cite{mcmillan2002measures, rezwanacofi}. A user's confidence in an AI agent's ability to perform tasks is improved when imbuing the agent with additional communication channels compared to the agent solely depending on conversation as the communication channel \cite{kim2018does}. 

In this paper, we investigate the impact of AI-to-human communication on the collaborative experience, user engagement and user perception of a co-creative AI. For the AI-to-human communication, we used speech, text and visual communication (an AI avatar). To guide the study design and data analysis, we formulated three research questions:

\begin{itemize}
\itemsep0em
\item {RQ1 - How does AI-to-human Communication affect the \emph{collaborative experience} in human-AI co-creation?}
\item {RQ2 - How does AI-to-human Communication affect \emph{user engagement} in human-AI co-creation?}
\item {RQ3 - How does AI-to-human Communication affect the \emph{user perception} of the co-creative AI agent?}
\end{itemize}

We developed two high-fidelity interactive prototypes of a co-creative system, Creative Penpal, that helps users in producing creative designs of a specific object by presenting inspiring sketches. One prototype utilizes only human-to-AI communication (baseline). The second prototype uses two-way communication between humans and AI, including AI-to-human communication via speech, text and a virtual AI avatar representing visual communication. We conducted a comparative user study with 38 participants to investigate the impact of including AI-to-human communication along with human-to-AI communication on collaborative experience, user perception and engagement. We present the findings as insights for making effective co-creative systems that will provide a better collaborative experience and increase user engagement. This paper makes the following contributions: 

\begin{itemize}
\itemsep0em
\item {We show that including AI-to-human communication improves the collaborative experience and user engagement compared to one-way human-to-AI communication in co-creative systems.}
\item {We present user perceptions of a co-creative AI with and without AI-to-human communication, highlighting the distinctions such as AI as a partner vs. tool.} 
\end{itemize}

This research leads to new insights about designing effective human-AI co-creative systems and lays a foundation for future studies regarding interaction design in human-AI co-creativity. 

\section{Related Work}
\subsection{Co-Creative Systems}
Research in computational creativity has led to different types of creative systems that can be categorized based on their purposes: 1) systems that generate creative products independently, 2) systems that support human creativity, and 3) systems that collaborate with users on a shared creative product as partners \cite{davis2015enactive}. In co-creative systems, humans and AI both contribute as creative colleagues in the creative process\cite{davis2013human}. Co-creative systems are supposed to be partners for humans and distinguished from autonomous creative systems, which generate creative products independently, and creativity support systems, which support human creativity \cite{kantosalo2020modalities}. Mixed initiative creative systems is another phrase that refers to co-creative systems \cite{yannakakis2014mixed}. Interaction is an intrinsic part of co-creative systems as both the human and the AI participate in the creative process and interaction makes the creative process complex and emergent. Maher explored issues related to \emph{who} is being creative when humans and AI collaborate in a co-creative system \cite{maher2012computational}. Antonios Liapis et al. argued that when creativity emerges from human-computer interaction, it cannot be credited either to the human or to the computer alone and surpasses both contributors' original intentions as novel ideas arise in the process \cite{liapis2014computational}. Designing interaction in co-creative systems has unique challenges due to the spontaneity of the interaction between the human and the AI \cite{davis2016empirically, kantosalo2014isolation}. A co-creative AI agent needs to be able to adapt to cope with human ideas and strategies. 
\vspace{-0.7cm}

\subsection{Interaction between humans and AI in Co-Creative Systems}
\textit{Interaction design} is the creation of a dialogue between users and the system \cite{kolko2010thoughts}. AI ability alone does not ensure a positive collaborative experience of users with the AI \cite{louie2020novice} and interaction is more critical than algorithms in the systems where interaction with the users is essential \cite{wegner1997interaction}. Bown asserted that the success of a creative system's collaborative role should be further investigated in terms of interaction design as interaction plays a vital role in the creative process of co-creative systems \cite{bown2015player}. Later Yee-King and d'Inverno argued for a stronger focus on the user experience, suggesting a need for further integration of interaction design practice into human-AI co-creativity research \cite{yee2016experience}. Kantosalo et al. said that interaction design, specifically, interaction modality should be the ground zero for designing co-creative systems \cite{kantosalo2020modalities}. Karimi et al. found an association between the way users interact in a human-AI collaboration and different kinds of design creativity \cite{karimi2020creative}. Tomaz and Chao showed that communication is an essential component in turn-taking based human-AI interaction \cite{thomaz2011turn}. 
\vspace{-0.2cm}  

\subsection{Communication in Human-AI Co-creation}
Communication is an essential component in any collaboration for the co-regulation between the collaborators and helps the AI agent make decisions in a creative process \cite{bown2020speculative}. A significant challenge in human-AI collaboration is the development of common ground for communication between humans and machines \cite{dafoe2021cooperative}. Previous work shows two-way communication between collaborators is essential in computer-mediated communication \cite{mcmillan2002measures}. AI-to-human communication represents the channels through which AI can communicate with humans, and this is essential in a human-AI co-creative system \cite{rezwanacofi}. AI-to-human communication is an essential aspect of human-computer interaction \cite{mcmillan2002measures}. In a co-creative setting, the modalities for AI-initiated communication can include text, voice, visuals (icons, image, animation), haptic and embodied communication \cite{nigay2004design}. Bente et al. reported that AI-to-human communication improved both social presence and interpersonal trust in remote collaboration settings with a high level of nonverbal activity \cite{bente2004social}. However, recent research has revealed that the majority of existing co-creative systems do not include AI-to-human communication, although it is critical in a human-AI collaboration for the AI to be considered as a partner rather than a tool \cite{rezwana2022designing}. Many co-creative systems include only human-to-AI communication through UI components like buttons/sliders, and the AI does not directly communicate to the human user \cite{rezwana2022designing}. For example, Image to Image \cite{isola2017image} is a co-creative system that converts a line drawing of a particular object from the user into a photo-realistic image: the user interface has includes a single button that users use to instruct the AI to convert the drawing. However, the AI cannot directly communicate with the user. 

Chatting with each other or using other types of communication channels increases engagement in a creative collaboration among humans \cite{bryan2012identifying}. Research shows that the way users talk in a human-AI conversation is similar to human-human conversation \cite{dev2020user}. Bown et al. explored the role of dialogue between the human and the user in co-creation and argued that both linguistic and non-linguistic dialogues of concepts and artifacts maintain the quality of co-creation \cite{bown2020speculative}. A recent study showed increased user satisfaction with text-based instructions rather than button-based instructions from the AI in a co-creation \cite{oh2018lead}. A user's confidence in an AI agent is improved when imbuing the agent with embodied communication and social behaviors compared to a disembodied agent using conversation alone \cite{kim2018does}. Additionally, the literature asserts that visual communication through embodiment aids synchronization and coordination in improvisational human-computer co-creativity \cite{hoffman2011interactive}. 

Researchers have investigated user perceptions of AI in different domains \cite{ashktorab2021effects, oliver2019communication, tijunaitis2019virtuality}, since social perception of one’s partner in a collaborative space can impact the outcome of the collaboration. The perceived interactivity – or lack thereof – of systems can have an impact on user perceptions of the system \cite{tijunaitis2019virtuality} as most existing co-creative systems use one-way communication. We build on the related research and the research gaps in existing co-creative systems' interaction designs to investigate the influence of two-way communication including AI-to-human communication.
\vspace{-0.2cm}  

\section{Creative PenPal: A Co-Creative System for Design Ideation}
Creative PenPal is a co-creative system that presents sketches to inspire users while they sketch design ideas in response to a specified design task. We developed two prototypes for Creative PenPal, one with AI-to-human communication and one without. The visual design of the interface is inspired by an existing co-creative system for design ideation, Creative Sketching Partner (CSP) \cite{karimi2020creative}. The difference between the two prototypes: one uses one-way communication, human-to-AI communication only (baseline), and the other has two-way communication, including AI-to-human communication. For the study reported in this paper, we did not implement the back-end AI since our research questions focus on the influence of communication in the interaction design. Therefore, we wanted to control the AI ability (same AI ability) in both versions so the study results are based only on the affect of AI-to-human communication. 

The prototypes for Creative PenPal offer three different kinds of design inspirations for the users - (a) sketches of the conventional design task object, (b) sketches of visually similar objects and (c) sketches of conceptually similar objects. Visually similar objects have visual or structural similarities to the user's sketch, and conceptually similar inspirations have similar themes or concepts as the design task object. For instance, when the design task object is a 'chair for gamers', Creative Penpal provides (a) sketches of typical chairs for gamers, (b) sketches of visually similar objects based on the user's sketch, which might be a wheelchair, ottoman, sofa, and (c) sketches of objects that are conceptually related to a chair for gamers, such as a keyboard, table, neck pillow, headphones. We selected a collection of sketches as the database. The sketches are grouped into three categories in the database based on the three kinds of sketches the system can present. We created a database of sketches for each of the two design tasks we used in the user study: a chair for gamers and a shopping cart for the elderly. 

The system randomly selects a sketch from the corresponding collection of sketches for conceptually similar object sketches and design-task object sketches. However, for the visually similar sketches, we used the Wizard of Oz (WOz) method to present sketches similar to users' drawings as a proxy for the AI. We used the WOz for visually similar object sketches as they need to be similar to what is being drawn on the canvas by users, unlike conceptually similar sketches that can be determined based on the design-task. In the user study, the Wizard could see the sketch on the User's Canvas and select a visually similar sketch to what was drawn by participants to display on PenPal's Canvas when participants clicked on the 'visually similar objects' button. The participants were unaware of the Wizard observing their sketch and were told that they were interacting with an AI. The visually similar object folder had 25 sketches for both design task objects, and the Wizard chose the most visually similar sketch to the participant's sketch. The same person was the Wizard for all study sessions to keep the methodology consistent. 
\vspace{-0.16cm}  

\subsection{Creative PenPal Prototype Without AI-to-human Communication (One-way Communication)}
\begin{figure*}[t]
  \centering
  \includegraphics[width=0.95\linewidth]{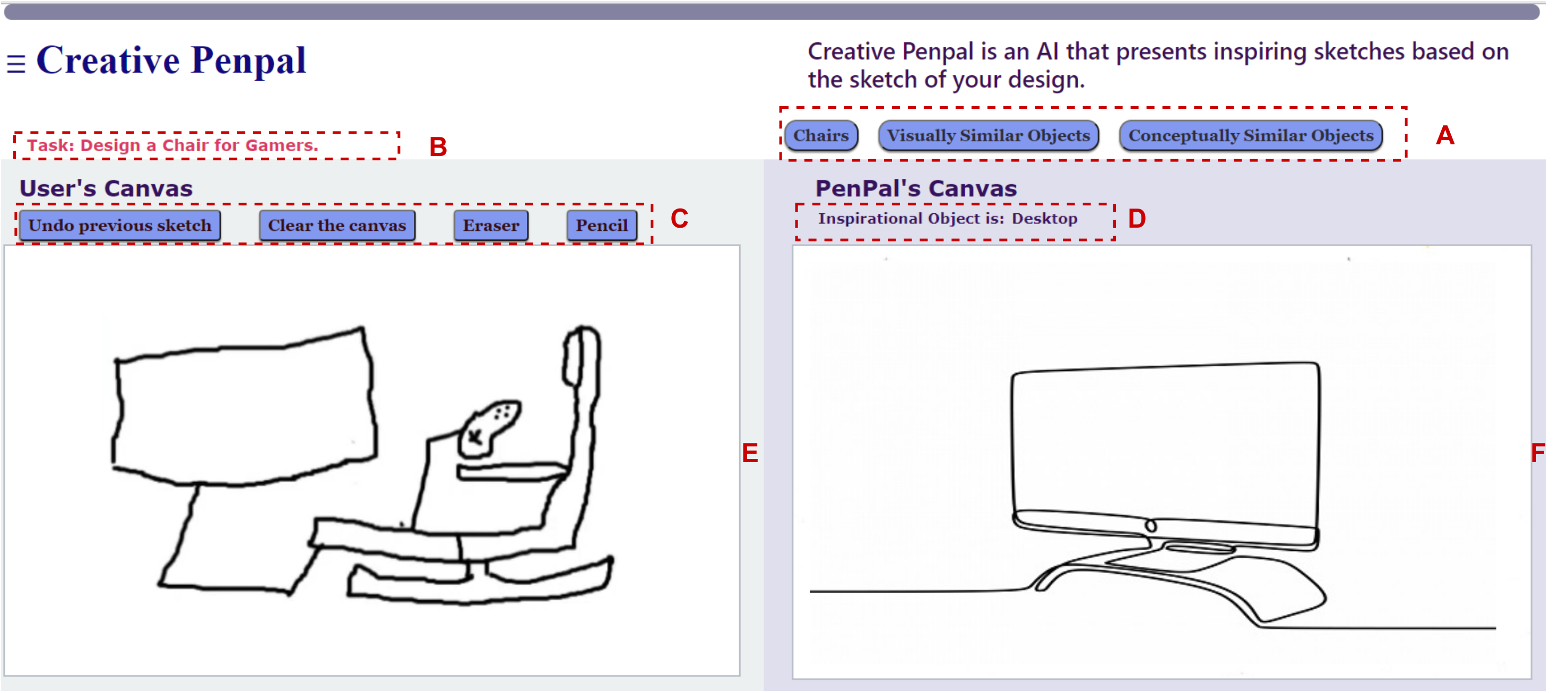}
  \caption{Creative PenPal Prototype without AI-to-human Communication (one-way communication): On the Left, Design Idea of a Participant. On the Right, AI showing an inspiring sketch}
  \Description{}
    \vspace{-0.6cm}  
  \label{instructing}
\end{figure*}

\vspace{-0.1cm}  
The baseline prototype, shown in Figure \ref{instructing}, uses buttons for human-to-AI communication to ask for different inspirations. The design task is shown on the interface in Label B (design a chair for gamers). Users design the object by drawing on the canvas depicted in Label E. Users can undo the last stroke using the button 'Undo Previous Sketch', erase a part of the sketch using the 'Erase' and erase the whole canvas by using the 'Clear the canvas' button (Label C). The 'Pencil' button is used to go back to the drawing (Label C). Users can ask for AI inspirations by clicking any of the three buttons in Label A. When users receive an inspiring sketch from the AI, they can see the sketch in the PenPal's Canvas (Label F) and the name of the inspiring object in the sketch shown in Label D.
\vspace{-0.16cm}  

\subsection{Creative PenPal Prototype With AI-to-human Communication (Two-way Communication)}
\begin{figure*}[t]
  \centering
  \includegraphics[width=0.95\linewidth]{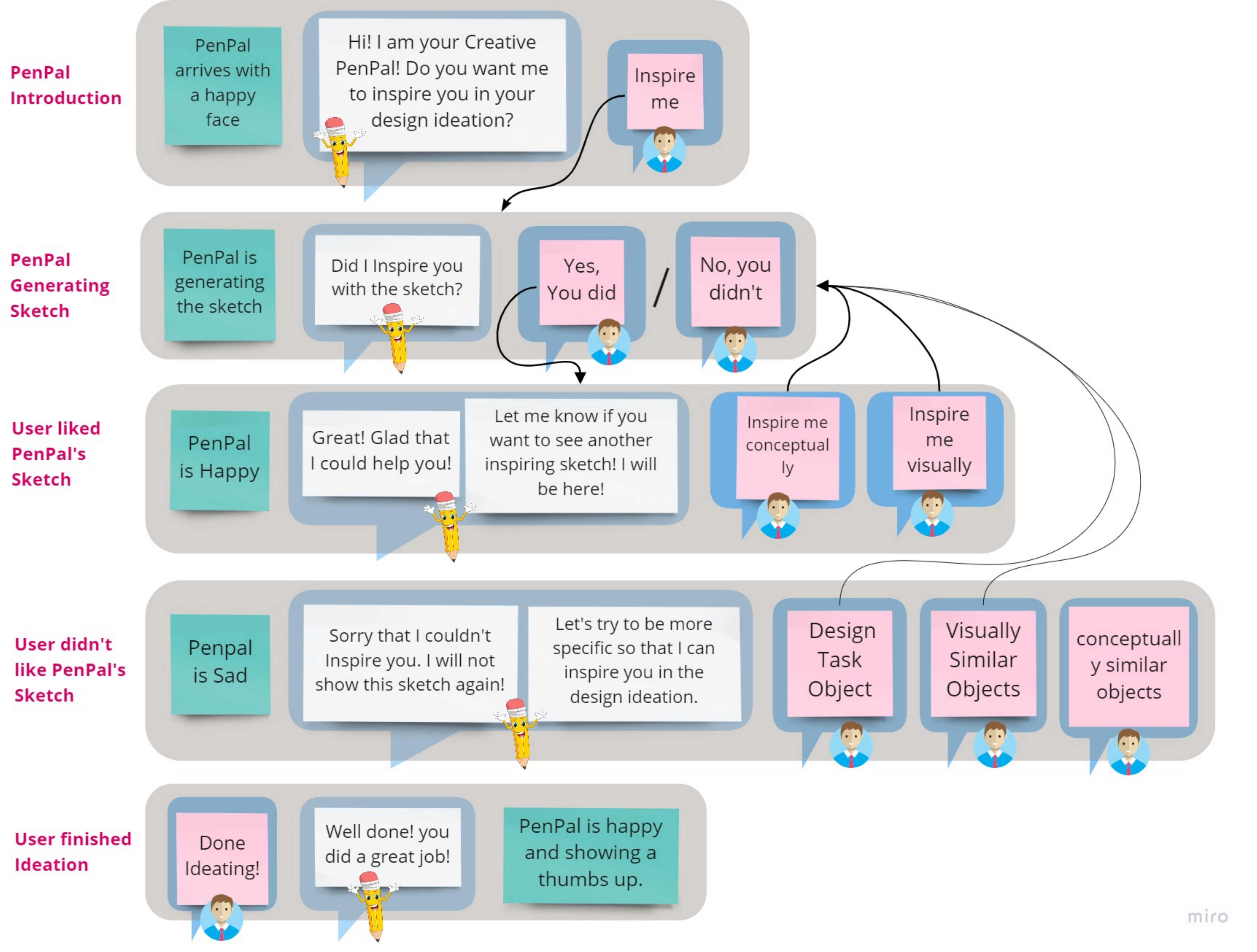}
  \caption{Two-way Communication Model including AI-to-human Communication for Creative PenPal}
  \Description{}
  \label{inter_model}
\end{figure*}

The prototype shown in Figure \ref{conversing} uses two-way communication between the human and AI. This prototype has the same human-to-AI communication as the baseline condition and uses AI-to-human communication through text, speech and visuals (an AI avatar). The AI avatar, a pencil (PenPal), is shown in Label G. Label A is where the AI communicates to the user via text, speech and the virtual AI avatar. The AI speaks the exact words as shown in the text. The AI voice is a recorded human voice that has been filtered through a robot-voice filter using free voice-altering software. When users click the 'Inspire me' button in Label A, the AI will show an inspirational sketch on its canvas in Label F. The users can also ask for three different kinds of inspirations using three buttons similar to the baseline prototype. The design task for this prototype is shown inside Label B (Design a shopping cart for the elderly). The human-AI conversational communication model for this prototype is demonstrated in Figure \ref{inter_model}. The communication model shows how two-way communication happens in 5 phases:

\begin{figure*}[t]
  \centering
  \includegraphics[width=1\linewidth]{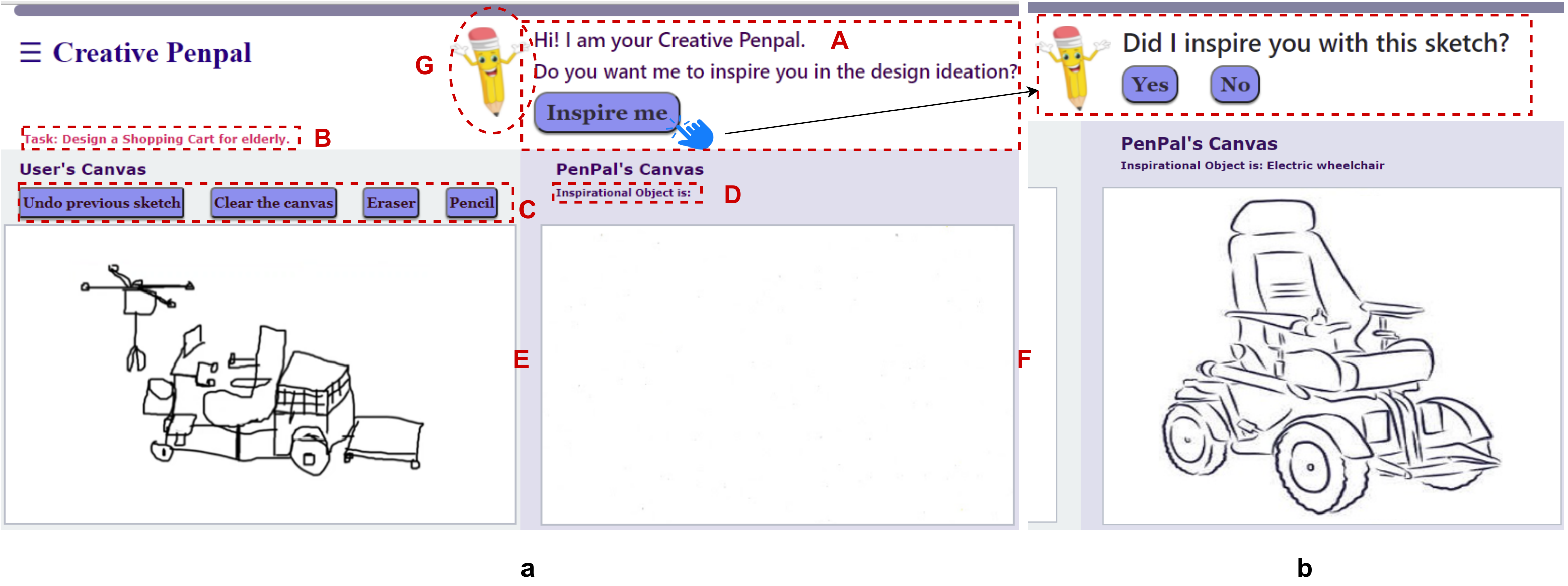}
  \caption{Creative PenPal prototype with AI-to-human communication (Two-way Communication): (a) PenPal introducing itself in the beginning of design ideation (b) User clicks on the 'Inspire me' button and PenPal shows a design inspiration}
  \vspace{-0.3cm}  
  \Description{}
      \vspace{-0.4cm}  
  \label{conversing}
\end{figure*}

\textbf{PenPal Introduction:} As soon as the user clicks on the start button to start the design task, the AI avatar arrives, introduces itself and asks the users if they want to see an inspirational sketch from the AI by saying, "Hi! I am your Creative PenPal. Do you want me to inspire you?". Users can respond immediately to get inspiration by pressing the button "Inspire me" or keep sketching to respond later. 

\textbf{PenPal Generating Sketch and Collecting User Preferences:} When the user hits the button "Inspire me", an animation of PenPal (the AI avatar) generating the sketch on the canvas is presented. After presenting an inspiring sketch, PenPal collects user preference by asking the user whether they liked the sketch or not. The user can reply with the "Yes" or the "No" button. 

\textbf{User Liked PenPal's Sketch:} When a user clicks the "Yes" button in response to PenPal's query about their preference, the PenPal arrives with a happy face and says, "I am glad that you liked the sketch! Let me know if you want another inspiration". If users want to see an inspiration again, they can click on the "Inspire me conceptually" or "Inspire me visually" button.

\textbf{User Did Not Like PenPal's Sketch:} When users click the "No" button, indicating that PenPal's sketch did not inspire them, PenPal arrives with a sad face and says, "Sorry that I could not inspire you! I will not show you this sketch again". Then it suggests: "Let's try to be more specific about what you want me to inspire with". The user can respond with any options, "Design Task Objects" (as our design task object is a shopping cart, the button says "Shopping Carts"), "visually similar objects", or "conceptually similar objects".

\textbf{User Finished Sketching:} The user finishes the design ideation task by clicking the "Finish Design" button. The virtual agent responds with: "Well done! You did a great job!"

\section{Comparative Study}
We conducted a comparative user study to explore the influence of AI-to-human communication on \emph{collaborative experience, user engagement} and \emph{user perception} of a co-creative AI using the prototypes. In this section, we describe the methodology of the study, the participants, the data, and the analysis.

\subsection{Study Methodology}
We used a within-subject approach for the study to collect quantitative and qualitative data to investigate our research questions. To see if condition order affects the outcomes, we counterbalanced the order of the conditions: half of the participants conducted their first design task with the baseline while the other half finished their first design task with the prototype featuring AI-to-human communication. 

\begin{figure*}[t]
  \centering
  \includegraphics[width=1\linewidth]{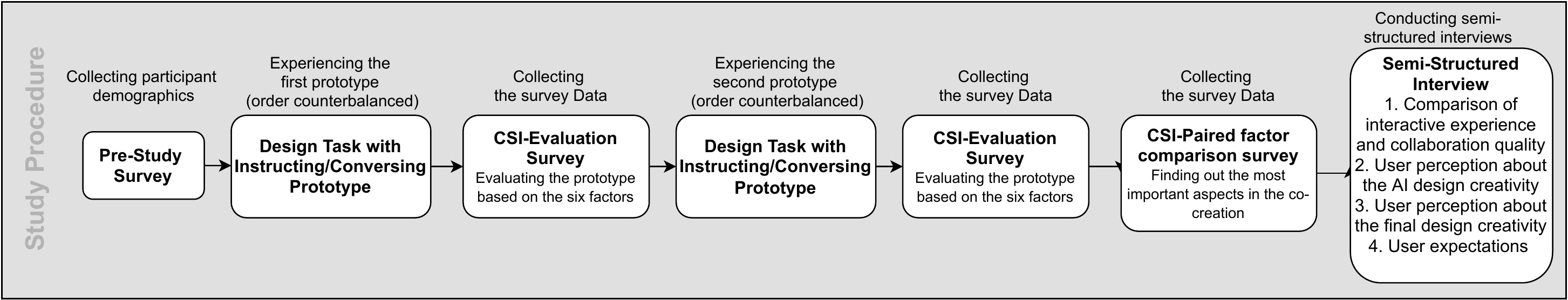}
  \caption{Study Procedure}
  \label{study_proc}
  \Description{}
  \vspace{-0.3cm}  
\end{figure*}

The study procedure is summarized in (Figure: \ref{study_proc}). At the beginning of the study, we collected demographic information from participants such as age, gender, and drawing/sketching ability using a survey. We briefly informed the participants about the design tasks while showing them the system interface before testing the prototypes. The design tasks were - 'Design a futuristic shopping cart for the elderly/a chair for gamers. Include at least three inspirations from the AI in your design idea'. We chose the design-task objects from everyday things of simple and similar complexity. Pilot studies showed no significant influence of the choice of design-task objects on the outcome. Participants shared their screens so that the wizard could see their designs. After each design task, the participants completed a survey to evaluate the system. Additionally, after completing both design tasks, they completed another survey to reflect on important aspects of their co-creation experience. Finally, the study ended with a follow-up semi-structured interview to collect qualitative data about the overall experience with the AI. 
    \vspace{-0.2cm}

\subsection{Participants}
We recruited 38 participants, 19 males and 19 females, who were all 18 years old or older (avg age = 26 years). We emailed participants an IRB-approved informed consent form to review and sign electronically upon scheduling the study. All participants voluntarily took part in the experiment and each participant received a gift card as an incentive upon completion of the study. The study did not require participants to have drawing/sketching skills. Among the participants, 23 participants had none/very little drawing/sketching skill, 14 participants had an intermediate skill of drawing/sketching, and 1 participant was an expert. 
    \vspace{-0.2cm}  

\subsection{Data Collection}

\subsubsection{Surveys}
In order to measure the perceived user engagement and overall experience with each prototype, we used Creativity Support Index (CSI) \cite{cherry2014quantifying}, a psychometric survey for measuring six factors in a creative system: \emph{Exploration, Expressiveness, Immersion, Enjoyment, Results Worth Effort} and \emph{Collaboration}. CSI consists of two separate surveys. The first survey evaluates a system using the six factors. For each factor, there are two agreement statements (12 statements in total). Participants rated each statement on a 10-point Likert scale of "Highly Disagree" to "Highly Agree". In the other survey, each factor is paired against every other factor (15 comparisons), which participants completed after finishing both design tasks. The latter survey is for determining the most important aspects of creative collaboration. The CSI survey is designed specifically for creativity support tools (CST) and the original collaboration factor is about human-human collaboration. Co-creative systems are distinct from CSTs as they are about human-AI collaboration. Therefore, we modified the original two agreement statements for the 'collaboration' factor to be more appropriate for evaluating human-AI collaboration: (1) The collaboration with the AI was more like interacting with a partner than a tool, (2) There was good and meaningful communication between me and the AI. The original statements for the CSI collaboration factor were: (1) The system allowed other people to work with me easily, and (2) It was really easy to share ideas and designs with other people inside this system \cite{cherry2014quantifying}.

\subsubsection{Interviews}
We collected in-depth qualitative data using semi-structured interviews. In the interviews, we questioned participants about (1) their interactive and collaborative experience with both prototypes, (2) their perceptions and satisfaction with the final designs created with both of the prototypes, (3) their perception of the co-creative AI in both prototypes and (4) their suggestions for improving their experience. During the interviews, we asked follow-up questions to dig deep and clarify interesting discussion points that came up in the conversation.
\vspace{-0.15cm}  

\subsection{Data Analysis}
\subsubsection{Surveys}
We conducted a  statistical analysis of the CSI survey data. We calculated the means and standard deviations for each factor score in the CSI and the final CSI score for both prototypes. We used T-tests comparing the effect of each condition on our outcome variables. We did not find any influence of study order, gender, age and drawing skill (independent variables) on any outcome variables (T-test and ANOVA). The P values are the following: study order (T-test, P = 0.3 for immersion, P = 0.24 for enjoyment, P = 0.05 for collaboration), gender (T-test, P = 0.08 for immersion, P = 0.46 for enjoyment, P = 0.5 for collaboration), age (Anova, P = 0.57 for immersion, P = 0.74 for enjoyment, P = 0.53 for collaboration) and drawing skill (Anova, P = 0.81 for immersion, P = 0.59 for enjoyment, P = 0.28 for collaboration).
\vspace{-0.15cm}  

\subsubsection{Interviews}
We conducted a thematic analysis of the interview data. As per Braun and Clarke's \cite{braun2012thematic} six-phase structure, two persons in the research team familiarized themselves with the interviews and created the initial codebook. The first author coded the interviews using the initial codebook (allowing for additional codes to develop). Following the coding process, both coders agreed on the codes to construct the primary themes.

\section{Results}
\subsection{CSI Survey Results}
A single CSI score is produced out of 100 for each prototype from the surveys. The average CSI score for communicating AI (condition B) is 80.95 (SD=11.90) and the score for baseline AI (condition A) is 73.096 (SD=16.671) (Table \ref{label1}). The T-test reveals a significant difference between the CSI scores of the two prototypes (P=0.021<0.05), indicating that participants rated the two versions substantially differently and rated the communicating AI higher than the baseline AI.

\begin{table}[]
\begin{tabular}{|c|c|c|}
\hline
{\textbf{Versions}} & \textbf{Average CSI Score (SD)} & \textbf{P Value}        \\ \hline
Communicating AI                                                       & 80.938 (11.898)                 &                         \\ \cline{1-2}
Baseline AI                                                      & 73.096 (16.671)                 & \multirow{-2}{*}{0.021} \\ \hline
\end{tabular}
\caption{Average CSI Score for both prototypes}
\label{label1}
\vspace{-0.3cm}  
\end{table}

\begin{figure*}[t]
  \centering
  \includegraphics[width=0.92\linewidth]{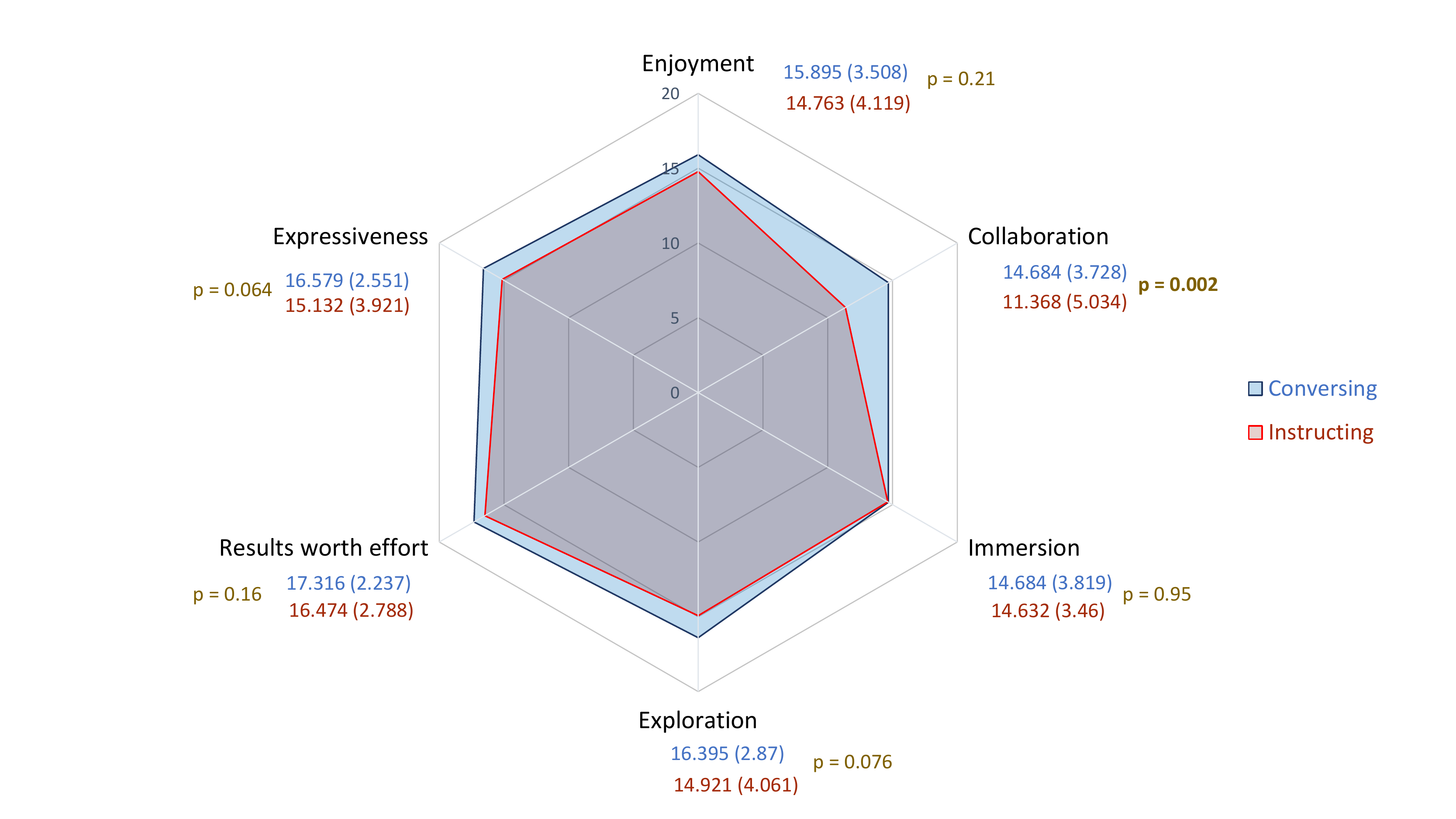}
      \vspace{-0.5cm}  
  \caption{Average scores for each CSI factor in both prototypes with standard deviation (inside brackets) and statistical significance (p)}
  \Description{}
  \label{radar}
\end{figure*}

Figure \ref{radar} shows the average scores for the six CSI factors for both prototypes. Participants scored each factor using a 20-point scale. We calculated the average scores for the six factors for each prototype and used a T-test to check their significance. All of the average factor scores for communicating AI are higher than the scores for the baseline. However, none of the factor scores significantly differ between the prototypes except the collaboration score. The average collaboration score for communicating AI is 14.684 and 11.368 for the baseline. The p-value from the T-test for the average collaboration score is p=0.002<0.05, which means participants scored the collaboration in communicating AI significantly higher than the baseline. Table \ref{label3} demonstrates that participants rated both statements of collaboration (sub-factors) for communicating AI significantly higher than the baseline. 

\begin{table}[]
\centering \scalebox{0.82}{
\begin{tabular}{|l|c|c|c|}
\hline
\textbf{Average collaboration factor agreements}                                  & \textbf{Communicating AI (SD)} & \textbf{Baseline AI (SD)} & \textbf{P Value} \\ \hline
There was good communication between me and the AI                                & 8.26 (1.826)             & 6.34 (2.581)              & \textbf{0.000}            \\ \hline
The interaction with the AI was "interacting with a partner" rather than a "tool" & 6.42 (2.41)              & 5.03 (2.8)                & \textbf{0.023}            \\ \hline
\end{tabular}}
\caption{Average scores for both agreements statements of Collaboration factor for both prototypes}
 
\label{label3}
 \vspace{-0.7cm}  
\end{table}

The paired factor comparison survey (the second survey of the CSI) results demonstrate the most important aspects of the co-creation to the participants. The most critical factors about the co-creation to the participants are exploration, result worth effort, following both collaboration and enjoyment (same average score). 

\subsection{Interview Results}
This section discusses the qualitative results and themes found from the thematic analysis. Two kinds of themes were found - some themes emerged directly from interview questions (high frequency) and some emerged from participants' unsolicited remarks and comments (low frequency). Results show that communicating AI was favored by a majority of the participants (n=26), the baseline AI was favored by 11 participants and 1 participant did not have any preference. The themes are described in the following subsections.

\subsubsection{Collaborative Experience (RQ1)}
\hfill \newline
\textbf{“It felt like working with someone”- communicating AI is perceived as a Collaborative Partner:} This was the most common theme that emerged from the interview. Most participants (n=30) stated that communicating AI provided a better collaborative experience since it seemed more like a collaborative partner than baseline AI. Participants described that communicating AI felt like a collaborative partner to them as it talked to them (n=24). P11 explained the interactive experience with the communicating AI saying, “\emph{The fact that it spoke with me was almost like an interaction with a person. I said, we don't need this idea, and it said I am not going to show you this picture again- which is more like working with a partner.}” Participants emphasized the human-like conversation characteristics of the AI as P21 said “\emph{The AI said, ‘I am so sorry. I will try again’. It gave me a feeling that I was interacting with somebody.}” P10 said, “\emph{The AI that spoke to me felt more like a partner compared to the other one where you would just click the buttons}”. On the same note, P1 expressed their preference that the AI talked to them by saying, “\emph{because it was talking to me and had the voice made it seem more like a human just saying sorry.}”

AI asking for human feedback (n=11) and the AI avatar (n=5) were also highlighted as factors for a better collaborative experience in addition to verbal communication. The feedback collection from the participants was compared to humans listening to their colleagues. Elaborating on this experience with communicating AI, P36 said, “\emph{It felt more human. It felt like you were actually working with a partner because you were actually getting and giving feedback back.}” p20 said, “\emph{It felt like a partner as it was allowing me to provide feedback}”. Participants mentioned that the affective characteristics of the AI avatar made them think of the AI as a partner. P2 said, “\emph{I liked it as it was happy that it helped me.}” P9 explained, “\emph{The friendliness and initial capturing of it by introducing itself made the system seem like it was meant to help you more than the other AI}(baseline)” P3 also reported similar experience when he said, “\emph{As soon as I went to the page and it asked me, 'hey, do you need some help?' That was really nice.}” 
\vspace{-0.3cm}  

\subsubsection{User Engagement (RQ2)}
\hfill \newline
\textbf{“I was more engaged with the system”- Increased user engagement with communicating AI:} Participants reported increased engagement, attention and enjoyment with the communicating AI, but none said anything about increased engagement or attention with baseline AI. Increased attention resulted from the awareness (n=10, unsolicited remarks) of another presence in the collaboration. Describing this awareness of being in collaboration with another entity, P9 stated, “\emph{The fact that it spoke to me really gave it a sense of its being.}” Participants also reported that the perceived co-presence aided them in developing a better design. For example, P12 said, “\emph{I felt like I was paying more attention to that system, maybe because someone was talking to me. I think that made my design a little bit better because my brain was thinking a little more.}” Participants were so engrossed in the collaborative experience that some of them desired to speak back to the AI. Like P17 said, “\emph{At some point, ithe AI actually made me speak back to it..}” Communicating AI increased user engagement and the creative potential of the final creative product compared to the baseline AI. P14 reported, “\emph{So, comparatively speaking, the second experience} (communicating AI) \emph{ was more engaging as in it felt like if I stayed and kept on drawing, I probably would have gotten more out of that experience.}” 

Participants also enjoyed collaborating with the communicating AI (n=26) more than the baseline A as it communicated back. P17 explained his experience and stated, “\emph{I feel like the first one}(communicating AI)\emph{ was more enjoyable to use. Like I, I actually had more fun doing it.}” P11 went on explaining the enjoyable experience, “\emph{The first one}(communicating AI)\emph{ was more enjoyable where it had the little pencil character and hearing it was more enjoyable and maybe a little bit easier.}” P5 described the fun part of interacting with the AI by saying, “\emph{In the second one}(communicating AI),\emph{ the interaction was more fun, as in like …it talked to you. It had like little animation. So that part was more I was more kind of invested in the second one.}”

\subsubsection{User Perception about Co-creative AI (RQ3)}
\hfill \newline
\textbf{“This AI is smarter and more helpful”- Users Perceived the communicating AI as Smarter AI:} Participants perceived the communicating AI as smarter than the baseline. P11 expressed their perception about the communicating AI saying, “\emph{Compared to the other one, I felt like the technology seemed a lot more advanced.}” P17 said on the same context that “\emph{I would say that the first version felt like AI} (communicating AI). “\emph{In the second version, I almost didn't even realize it was an AI.}” Other participants felt like the communicating AI understood their needs better than the baseline and was in sync with their thoughts. “\emph{The first one} (communicating AI)\emph{ was more in sync with my thoughts and was more AI-ish}”, said P4.

Participants also thought that the communicating AI was more helpful and reliable in guiding them through the creative process than the baseline AI. For example, P7 said, “\emph{The second one} (communicating AI)\emph{ was definitely, more helpful in allowing me and in guiding me to what I wanted to draw.}” P23 elaborated on how the communicating AI helped in being more creative, “\emph{It was really helpful because if you had just asked me to draw a shopping cart without any AI, you would probably see a shopping cart that you see at any normal Wal-Mart or Target. I feel like it wouldn't have been as creative. It really did help with the creativity and it was really beneficial.}” P7 shared how the communicating AI was more reliable, “\emph{The second one kind of guided me to what I wanted to draw...I think I came to my conclusion quicker.}”

\textbf{“It was like searching images on Google”- User Perceptions about the baseline AI:} When describing the interactive experience with the baseline AI, participants compared the experience with a Google image search (n=6, unsolicited remarks). For example, P22 said, “\emph{It felt like searching images on Google or something to look at pictures.}” Comparing their experiences with baseline AI and the communicating AI, P16 said, “\emph{The first version wasn't as interactive} (baseline). “\emph{It was kind of the equivalent of looking up kind of images on Google, because it wasn't speaking to me.}” Participants explained how they felt the baseline AI was aloof and did not communicate with them by saying, “\emph{It's like when you go to Google and you search something, Google is not going to say, hey, thank you for searching this, and here are the results.}” P16 described their experience with baseline and said, “\emph{The first version }(baseline),\emph{ didn't really feel like it was any form of AI. Kind of felt like a photo refresher on Google.}”

Just users clicking buttons to give the AI basic instructions and no communication from the AI, led the participants to perceive it as a tool and not as an intelligent colleague. Elaborating on this experience, P22 said, “\emph{It didn't really feel like an AI. I just felt like something generating images.}” Some participants even reported that the baseline version felt like a random image generator. For example, P33 said, “\emph{The second one felt more like... let's throw things at the wall and see what sticks…the way they had at least. because like they didn't ask me for anything.}”

\textbf{“It felt a lot more Personal”- Personal Connection with communicating AI:} Participants reported that the communicating AI felt more personal and personable (n=5, unsolicited remarks). Because of its human-like attributes, such as verbal and visual communication, the AI was perceived as a persona that made it more intimate and connected. Regarding this P12 said, “\emph{I think just adding the simple feature, like speaking to you and listening to you, made it more personal. The AI was like, oh, that didn't work! Let's try something else. It just made the AI more personal that I would be more likely to use.}” P24 described how the feedback collection made them feel more included in the collaboration by saying, “\emph{It would take the feedback that you give it and change the images that it gave you based on that. So it was definitely a lot more personal.}”

Participants also spoke about how they and the AI had a mutual understanding, which made the AI feel personal. In this context, P36 elaborated on their experience, “\emph{It gives me an idea and ask, did you like the idea? And I'm like, yeah, I like the idea or no, the idea is bad but maybe we can incorporate this. So the second one was a lot more personal.}” On the same context, P33 said, “\emph{Well, the first version looks more like personal where it was asking questions and all that felt more like a partner than the other one.}”

\hfill \newline
\textbf{“My final design is more creative where the AI talks” – Perceived Creativity with communicating AI vs baseline AI:} In response to the interview question about the final designs, most participants (n=27) expressed their satisfaction with the design created with the communicating AI. P9 elaborated on this and said, “\emph{I felt like I developed a better final product and to me, using AI is about coming out with a very efficient design. And I felt that the second one} (communicating AI)\emph{ was able to make it that way.}” Many participants thought the final design with the communicating AI had more potential to be a good design. For example, P14 reported, “\emph{I like the second design} (communicating AI)\emph{ because I think it has more potential. So if you continue to work on it with the assistance of the AI, I think it has more potential.}” 

Participants explained that they thought the communicating AI reassured them about the inspirations it would show them, so they felt confident about their final design. P32 said, “\emph{After collecting feedback, it reassured me that it would not show the same sketch twice for our design compared to the first.}” The intriguing aspect of this remark is that the participant thought it was 'their' idea rather than 'his', indicating a sense of the shared creative product. Participants also felt included in the final design created with the communicating AI, unlike the baseline AI. P36 reported regarding this issue and said, “\emph{I think that this design} (communicating AI)\emph{ is a lot more influenced by AI because with the first design} (baseline) \emph{I was not really included.}”
 \vspace{-0.25cm}  
 
\subsubsection{User Expectations (Additional Findings)}
\hfill \newline
\textbf{User wants the AI to Speak like a Human:} Participants expect a human-like friendly voice of a co-creative AI partner in a conversing interaction. Even though the voice of the communicating AI was a recorded human voice, we used a free voice changer app, so it has a slightly robotic tone. Some participants did not like that voice. For example, P16 said, “\emph{The voice was a little creepy and distorted…if you've ever played for five nights at Freddie, a video game, it's what I would imagine one of those like horror robots to sound.}” They advised changing the voice to be more human-like and pleasant in order to appear more welcoming and collaborative. Like P22 said, “\emph{I think the only thing is with the second version}(communicating AI)\emph{...making it talk more like human-like.}” Participants advised that the AI not use the exact phrase everytime it delivers the same information and rather use alternative phrases to convey the same thing. About this, P22 said, “\emph{It would repeat the same thing every time. I guess it has to say the same thing each time, but maybe it should do that with different phrases.}” Some participants suggested implementing two-way voice communication. For example, P19 said,“\emph{It would've been cool if I could like talk to the AI. Like, can you show me the next one?}”

\textbf{Users Want Flexibility over using the AI Contributions:} Participants want flexibility over how they can use the AI contributions. Some participants (n=4) suggested that options should be available to go through the inspirations previously shown by the AI. They reported that they realized the value of certain inspirations only after they were gone and new inspirations had been shown to them. P34 suggested saving a list of previously shown image inspirations by the AI and said, “\emph{If I can go forward to get back to the previous pictures that I have already seen, it would have been better.}” P35 proposed a 'maybe' button, which would display the inspirations that participants believed could be helpful for later usage and suggested, “\emph{Add another button that will say ‘maybe’, to store inspiring sketches that I might use later.}”
 \vspace{-0.25cm}  

\section{Discussion}
In this section, we begin by revisiting our research questions with a summary of our findings, followed by a discussion of the implications of the findings on designing human-AI interaction in co-creative systems that lead to a more engaging collaborative experience.  

\textbf{RQ1 - How does AI-to-human Communication affect the collaborative experience in human-AI co-creation?} Our results show that two-way communication, including AI-to-human communication, improves the collaborative experience in human-AI co-creation. The survey results showed that participants scored their collaborative experience with communicating AI significantly higher than the baseline. Interviews revealed that most participants reported their experience with the communicating AI as more like collaborating with a partner, unlike the baseline AI. Most participants liked the aspect that the AI spoke to them and collected feedback from them. Participants also liked the affective characteristics of the AI character displays, like visually being sad when users did not like its inspirations or visually being happy when they liked an inspiration.

\textbf{RQ2 - How does AI-to-human Communication affect user engagement in human-AI co-creation?} The results from thematic analysis demonstrated that participants engaged more with the communicating AI than with the baseline. Most participants enjoyed using the communicating AI more than the baseline AI. Participants also reported being in sync with the communicating AI and wanted to talk back to the AI but not with the baseline AI. Participants reported a sense of awareness of another ‘being’ during collaboration with the communicating AI, which helped them be attentive and engaged.

\textbf{RQ3 - How does AI-to-human Communication affect the user perception of the co-creative AI agent?} The communicating AI was perceived as the smarter and more reliable AI. Many participants perceived that communicating AI helped and guided them more than baseline AI. Many participants compared the experience with the baseline as a Google image search. Participants also perceived the communicating AI as more personal as they felt connected with it. Most participants preferred the final design created with the communicating AI as more creative. Additionally, participants expect the AI to communicate with them more like a human than a robot. 

Participants expressed additional interaction design features to improve the human-AI collaboration. Participants wanted flexibility over how they could use the contributions from the AI. Some participants mentioned that efficiency and interaction time mattered to them, and they wanted the interaction to be faster. Most participants who preferred the baseline AI (n=9 out of 11) liked it because interaction was faster with only clicking buttons compared to the communicating AI, even though they thought the communicating AI was more partner-like and engaging. This reveals the importance of the efficiency of the communication between users and AI. Efficiency is one of the most critical factors for user experience and further research should be done to design two-way communication more efficient. Additionally, some participants suggested less frequent feedback collection as our communicating AI asked the user for their preference every time it showed an inspiration. Another user expectation is diverse contributions from the co-creative AI as they thought that diverse contributions would produce a more creative shared product in the co-creation.

With advances in AI ability in human-AI co-creative systems, there is a need for human-centered research focusing on user engagement and successful collaborative experiences. Unlike autonomous generative AI, a fundamental property of co-creative systems is the interaction between humans and AI as partners. Therefore, advances in interaction design along with AI ability are needed. Since user perception of one’s partner in a collaborative space can impact the outcome of the collaboration, user perception of AI is an important consideration. As technology advances, the perception of AI and expectations from AI change. People use commercial conversational AI like Siri and Alexa every day and people are now familiar with AI that talks to them. These conversational AI set the norm of AI talking and communicating with people. “\emph{Did you see the movie Iron man? It was like Jarvis helping me}”, said P8, who expressed satisfaction with communicating AI as it matched with her perception of advanced AI. Based on our findings, AI-to-human communication through voice and visuals can be implemented in co-creative AI to improve collaborative experience and engagement as they provide a sense of co-presence and partnership. However, two-way communication between humans and AI should meet current expectations, as we found that users don’t like repetitive dialogues, making it less human-like. Including affective characteristics in the speech and embodiment of the AI make it more personal-another insight our study revealed. Including affect can be used to increase the personalization of co-creative AI.

One-way communication(human-to-AI) might limit the engagement and enjoyment with the system. Clicking buttons without any other communication channel might change the perceived ability of the AI even though the algorithm is powerful. For example, if users think co-creative systems are just like a Google search, they may not see the value of the AI. People have acquired bias towards how they interact with AI versus humans. Prior research shows that in human-AI collaboration, when users perceive their partners to be human, they find them to be more intelligent and likable \cite{ashktorab2020human}- as one of our participants said “\emph{I would rather collaborate with a human.}" However, two-way communication, including AI-to-human communication, can make a significant difference as participants perceived communicating AI as the more reliable and smarter AI and the final product more creative. The two-way communication provoked a sense of reliability, like a P32 said, “\emph{It was more reassuring}.” Trust and reliability are essential in collaboration and our results showed that even if the ability of AI is the same, the communication style influences the way users trust and rely on a co-creative AI partner. Some participants wanted to talk back to the AI as it seemed more fun, personal and reliable. Our findings show that further research to identify ethical issues is needed as ethical issues may arise with users relying on the AI too much and revealing unintended data to the AI.

\vspace{-0.3cm} 
\section{Limitations}
Even though we chose simple and similar complexity design task objects, the exact design tasks chosen for each prototype may have an impact on the results. The AI-to-human communication model between the human and AI in our prototype comprises a set of simple predefined speech and texts. A more refined AI-to-human communication model can communicate with users using less repetition and more context-specific speech, text and visuals. A limitation of this study is whether the results with a basic two-way communication transfer to a more dynamic two-way communication between humans and AI. However, we expect a more refined communication model to improve the user experience. Secondly, our study is based on a high-fidelity prototype of a co-creative AI where a wizard (WOz) selected visually similar inspiring sketches based on users' drawings. A fully implemented AI may offer different sketches than a human wizard and inspire users differently. Additional studies are needed to examine the user experience with a refined two-way human-AI communication and a fully implemented AI. With these limitations, we see our findings as preliminary, indicating areas for future work. 

\vspace{-0.2cm} 
\section{Conclusion}
In this paper, we investigate the influence of two interaction designs, with and without AI-to-human communication, on \emph{collaborative experience, user engagement} and \emph{user perception} of the co-creative AI using a comparative study. We designed two prototypes for the study and identified that including AI-to-human communication along with human-to-AI communication improves the collaborative experience and user engagement as the co-creative AI is perceived as a collaborative partner. Including AI-to-human communication also positively changes user perception of co-creative AI as users perceive it as more intelligent and reliable. This research leads to new insights about designing effective human-AI co-creative systems and lays the foundation for future studies. Additionally, insights from this research can be transferred to other fields that involve human-AI interaction and collaboration, such as education, entertainment, and professional work.

\bibliographystyle{ACM-Reference-Format}
\bibliography{sample-base}


\end{document}